\documentclass[aps,prb,superscriptaddress,reprint]{revtex4-2}
\usepackage[T1]{fontenc}
\usepackage{newtxtext}
\usepackage{newtxmath}
\usepackage{amsmath}
\usepackage{physics2}
\usephysicsmodule{ab,ab.legacy,braket,nabla.legacy,op.legacy}
\usepackage{diffcoeff}
\usepackage{mathtools}
\usepackage{siunitx}
\usepackage{comment}
\usepackage[dvipdfmx]{graphicx}
\usepackage[dvipsnames]{xcolor}
\usepackage{hyperref}
\hypersetup{colorlinks=true,linkcolor=Blue,citecolor=Blue,urlcolor=Blue}
\usepackage{bm}

\begin{document}
\title{Description of molecular chirality and its analysis with high harmonic generation}
\author{Akihito Kato}
\affiliation{Department of Physics and Electronics, Osaka Metropolitan University, 1-1 Naka-ku, Sakai, Osaka 599-8531, Japan}
\author{Nobuhiko Yokoshi}
\affiliation{Department of Physics and Electronics, Osaka Metropolitan University, 1-1 Naka-ku, Sakai, Osaka 599-8531, Japan}
\date{\today}

\begin{abstract}
  To clarify the microscopic origin of chirality-induced optical effect, we develop an analytical method that extracts the chiral part of the Hamiltonian of molecular electronic states. We demonstrate this method in a model chiral molecule consisting of two helically stacked $N$-sided regular polygons, and compare it with numerical calculation for chiral discrimination via high-harmonic generation (HHG) of the same molecule.
  The discrimination signal here is the Kuhn $g$ factor, the difference between the harmonic intensity from the bicircular laser field and that from its reflected laser field normalized by their average. The $g$ factor is a pseudoscalar quantity that reflects the chirality of the molecule. As a result, we find that the $g$ factor becomes large over a wide range of harmonic orders making HHG suitable for chiral discrimination. We further find that, to increase the difference of harmonic intensity from the above two fields, the unnormalized $g$ factor, the increase of capacity to generate the longitudinal dipole moment is more advantageous than maximizing the transverse-to-longitudinal conversion efficiency via the optimization of molecular chirality.
  We speculate this criteria may be extended to other optical and current-induced processes relevant to the chiral molecules and materials.
\end{abstract}
\maketitle

\section{\label{sec:intro}Introduction}

Chirality, a structural property of an object
that lacks reflection and inversion symmetries,
is ubiquitous in nature from single molecules, molecular assemblies, and crystals, to living organisms
~\cite{wagniere2007chirality}.
Much effort has been devoted toward understanding the functionality inherent in the chiral systems.
In crystals,
electronic and phononic band splittings are originated from the chiral symmetry breaking
~\cite{bozovic1984Possible,ishito2022Truly,ishito2023Chiral,ueda2023Chiral,oishi2024Selective,tsunetsugu2023Theory,kato2023Note},
which has invigorated possibilities in transporting and converting their angular momentums (AMs)
~\cite{zhang2015Chiral,fransson2023Chiral,ohe2024ChiralityInduced}.
In particular,
properties of the AM in the chiral systems and its relevance to chirality-induced spin selectivity (CISS)
~\cite{gohler2011Spin,inui2020ChiralityInduced,evers2022Theory,bloom2024Chiral}
effect have been increasingly focused on.
Thus,
as well as elucidating the characteristics of chiral materials
and establishing efficient means of asymmetric synthesis,
discriminating the chirality is also the indispensable problem.

Recent studies have demonstrated that the chirality in quantum systems can be represented as the electric toroidal monopole (ETM)
~\cite{oiwa2022Rotation,kishine2022Definition,kusunose2024Emergencea,inda2024Quantificationa}.
This is anti-symmetric with respect to the inversion and reflection and has the time-reversal symmetry, and thus is a generalization of the definition of the chirality by Kelvin and Barron~\cite{barron1986Symmetry}.
Because also the Hamiltonian of chiral systems should include the ETM, we can formally separate it into the symmetric and anti-symmetric part.
The anti-symmetric part is defined by the subtraction of the Hamiltonians with the opposite chiralities, which belongs to the different Hilbert spaces composed of the different electronic basis set.
Then, its direct evaluation has a major difficulty.
One of the main results of this work is that we develop a method to enable the direct subtraction of the matrix elements, and thus the identification of the chirality.

Using the above identification method, we consider the chiral discrimination signal by light as one example.
The chiral systems asymmetrically interact with the chiral light, which can be utilized to optically discriminate the chirality.
In circular dichroism (CD),
the differential absorption spectrum between the clockwise and the counterclockwise circular polarization of light,
which correspond to the spin angular momentum (SAM), is measured.
The measured spectrum is the pseudoscalar quantity, i.e., each enantiomer has the opposite sign from the other.
The ability to extract the pseudoscalar (more generally the pseudotensor) quantity can be a clear manifestation of the chiral discrimination~\cite{ordonez2018Generalized,rouxel2022Molecular}.
It is natural to explore criteria from a microscopic point of view on how to extract a large pseudoscalar quantity.

Here, we focus on a nonlinear optical process called high harmonic generation (HHG), which is the light emission with frequencies multiple integers of the injected one.
As a consequence of the SAM conservation, the atoms and molecules cannot emit the high harmonics from the light beams with the single circular polarization and requires the bicircular field, the combination of both circular polarizing fields~\cite{eichmann1995Polarizationdependent,milosevic2015Circularly,huang2018Polarization}.
This bicircular field can break the reflection symmetry, and thus, can be chiral.
Then, the chiral discriminating HHGs using the bicircular field have been studied~\cite{cireasa2015Probinga,harada2018Circular,smirnova2015Opportunities,wang2017High,ayuso2018Chiral,neufeld2019Ultrasensitive,ayuso2022Strong},
wherein the HHG spectrum measurements have demonstrated that its Kuhn $g$ factor is pseudoscalar.
We demonstrate our developed identification method in a model chiral molecule consisting of two helically stacked $N$-sided regular polygons, and compare it with numerical calculation within so-called three-step model~\cite{corkum1993Plasma,lewenstein1994Theory}.
It is found that the increase of the longitudinal dipole moment is more advantageous than maximizing the transverse-to-longitudinal conversion efficiency to obtain the large difference between the harmonic intensity from the bicircular laser field and that from its reflected laser field.

The remaining part of this paper is organized as follows.
In Sec.~\ref{sec:model}, we introduce the model chiral molecule
and its anti-symmetric part is deduced in Sec.~\ref{sec:etm}.
In Sec.~\ref{sec:hhg}, we present the theory of the HHG based on the three-step model and prove that the Kuhn $g$ factor becomes pseudoscalar using the symmetry argument.
In Sec.~\ref{sec:result}, numerical results of the chiral discriminating HHGs are presented and its correspondence to the chiral part of the Hamiltonian are analyzed.
Section~\ref{sec:conclu} is devoted to the concluding remarks.

\section{\label{sec:model}Model Chiral Molecule}

\begin{figure}
  \centering
  \includegraphics[width=80mm]{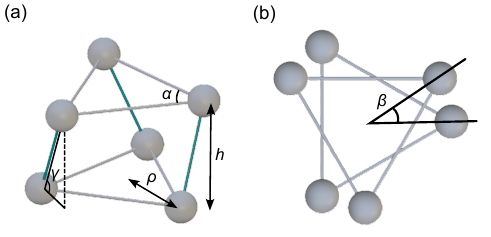}
  \caption{(a) Illustration of the model chiral molecule that consists of two $N=3$ regular polygons. The gray and green lines display the intralayer and interlayer bonds. (b) Top view of our model. Regular polygons are helically stacked with the twist angle $\beta$.}
  \label{fig:model}
\end{figure}

As a model chiral system,
we consider two helically stacked $N$-sided regular polygon molecules (see Fig.~\ref{fig:model} for $N=3$)
characterized by the radius $\rho$, the interlayer distance $h$, and the twist angle $\beta$.
The position of the $j$th atom ($j=1,\ldots,N$) in the $k$th layer ($k=1,2$) is written as
$\bm{R}_{kj}(\beta) = [\rho\cos2\theta_{kj}, \rho\sin2\theta_{kj}, (k-1)h ]^\top$
with $\theta_{kj} = (k-1)\beta/2 + (j-1)\alpha/2$ and $\alpha=2\pi/N$ being the rotation angle.
To break the reflection symmetry, the twist angle has to satisfy $0 < \vab{\beta} < \alpha/2$.
Chirality reversal corresponds to the inversion of the twist angle, $\beta \to -\beta$,
under the action of the reflection operator $\hat{\sigma}$ with respect to the $xz$ plane,
$\hat{\sigma}\bm{R}_{kj}(\beta)=\bm{R}_{kj}(-\beta)$.

Based on the Slater--Koster method~\cite{harrison1989electronic,varela2016Effective},
we construct the tight-binding Hamiltonian $\hat{H}(\beta)$ using the electronic basis states
$\{ \ket*{\phi_{kj}^\nu(\beta)}\}$
that consist of $2\mathrm{s}$ ($\nu=0$) , $2\mathrm{p}_z$ ($\nu=z$), and $2\mathrm{p}_\pm = 2\mathrm{p}_x \pm i2\mathrm{p}_y$ ($\nu=\pm$) orbitals from each atom,
where $\ket*{\phi_{kj}^\nu(\beta)}$ represents the orbital $\nu$ centered at $\bm{R}_{kj}(\beta)$,
$\braket*{\bm{r}}{\phi_{kj}^\nu(\beta)} = \phi^\nu[\bm{r}-\bm{R}_{kj}(\beta)]$.
We consider the electronic bonds between the adjacent atoms in the same layer (intralayer) and the ones with the same $j$ in the different layers (interlayer).
The latter bond is necessary to endow the chirality with the electronic state, which is characterized by the helical angle $\gamma = \arctan[h/2\rho\sin(\beta/2)]$ appeared in the form of $\cos\gamma$ and $\sin\gamma$. The former changes its sign under the chirality reversal, while the latter is unchanged. In the limit of $h/\rho \to \infty$ or $\beta \to \pm0$, the molecule becomes achiral as the former vanishes, $\cos\gamma \to 0$. For $\gamma = 0$ with $h = 0$, the molecule becomes the two-dimensional system that cannot be chiral~\cite{kusunose2024Emergencea}.
The Slater--Koster overlaps for the intralayer bond are denoted by $V_{\lambda\nu}^s$
($s=\sigma$ and $\pi$ for parallel with and perpendicular to the bonding),
which is parametrized by $V$,
and those for the interlayer bond are denoted by $W_{\lambda\nu}^s$,
which is parametrized by $W$.
For the constructed Hamiltonian and its detailed derivation, see Appendix.~\ref{sec:tb_detail}.

The groundstate $\ket{\phi_\mathrm{GS}(\beta)}$ is the eigenstate of $\hat{H}(\beta)$,
and hence, belong to the irreducible representation (IRREP) of the point group $\mathrm{C}_N$,
because $\hat{H}(\beta)$ possess the $N$-fold rotational symmetry.
Each eigenstate $\ket*{\phi_m(\beta)}$ is classified by its rotational quantum number or the pseudo AM~\cite{zhang2015Chiral,zhang2022Chiral}, $m$, through the relation
$\hat{C}_N \ket{\phi_m(\beta)} = e^{-im\alpha} \ket{\phi_m(\beta)}$
with $\hat{C}_N$ being the $N$-fold rotational operator.
For $N$-odd and even cases, the pseudo-AM takes the value $m = 0, \pm 1, \ldots, \pm (N-1)/2$ and $m=0, \pm1, \ldots, \pm(N-1)/2, N/2$, respectively.
The time-reversal symmetry enforces the degeneracy of $\ket*{\phi_m(\beta)}$ and $\ket*{\phi_{-m}(\beta)}$.
Thus, the ground state becomes $\ket{\phi_\mathrm{GS}(\beta)} = [\ket*{\phi_m(\beta)} + \ket*{\phi_{-m}(\beta)}]/\sqrt{2}$.

\section{\label{sec:etm}Description of Molecular Chirality}

To discuss the correspondence between the molecular chirality and the Kuhn $g$ factor,
we extract the anti-symmetric part of the Hamiltonian
with respect to the reflection operation.
This part is formally written as
$\hat{G}_0(\beta) = [\hat{H}(\beta)-\hat{H}(-\beta)]/2$,
which satisfies
$\hat{\sigma}\hat{G}_0(\beta)\hat{\sigma}^\dagger = -\hat{G}_0(\beta)$.
Because the matrix element
$\braket*[3]{\phi_{ki}^\lambda(\beta')}{\hat{H}(\beta)}{\phi_{lj}^\nu(\beta')}$
is well defined only for the $\beta=\beta'$ cases,
the direct subtraction of the Hamiltonians with the opposite $\beta$ requires the
transformation from $\ket*{\phi_{kj}^\nu(0)}$ to $\ket*{\phi_{kj}^\nu(\pm\beta)}$.
For that purpose,
we first translate the orbital center from $\bm{R}_{kj}(\beta)$ to $\bm{R}_{kj}(0)$ as
\begin{align}
  \phi^\nu[\bm{r}-\bm{R}_{kj}(0)]
   & = \exp\ab[-\frac{i}{\hbar}\bm{\hat{p}}\cdot[\bm{R}_{kj}(0)-\bm{R}_{kj}(\beta)] ]
  \notag                                                                              \\
   & \quad \times \phi^\nu[\bm{r}-\bm{R}_{kj}(\beta)]
  \notag                                                                              \\
   & = \hat{T}_{kj}(\beta) \phi^\nu[\bm{r}-\bm{R}_{kj}(\beta)],
\end{align}
where $\hat{\bm{p}}=-i\hbar\bm{\nabla}$ is the momentum operator.
This procedure is similar to pullback or push forward mapping in the Lie derivative~\cite{nakahara2003Geometry}.
Expressing the orbital as the product of the radial distribution and the spherical harmonics,
the achiral basis with $\beta=0$ can be written as a linear combination of the chiral basis states,
\begin{equation}
  \ket*{\phi_{kj}^\nu(0)}
  = \hat{T}_{kj}(\beta)\ket*{\phi_{kj}^\nu(\beta)}
  = \sum_{\nu'} g_{kj}^{\nu\nu'}(\beta) \ket*{\phi_{kj}^{\nu'}(\beta)}
\end{equation}
with $g_{kj}^{\nu\nu'}(\beta) \in \mathbb{C}$
(see Appendix.~\ref{sec:g0_detail} for details).
This allows us to expand the Hamiltonians with the achiral basis states as
$\hat{H}(\beta)
  = \sum_{klij\lambda\nu} \braket*[3]{\phi_{ki}^\lambda(0)}{\hat{H}(\beta)}{\phi_{lj}^\nu(0)}
  \ketbra*{\phi_{ki}^\lambda(0)}{\phi_{lj}^\nu(0)}$
with
\begin{align}
   & \braket*[3]{\phi_{ki}^\lambda(0)}{\hat{H}(\beta)}{\phi_{lj}^\nu(0)}
  \notag                                                                 \\
   & = \sum_{\lambda'\nu'}
  (g_{ki}^{\lambda\lambda'}(\beta))^\ast g_{lj}^{\nu\nu'}(\beta)
  \braket*[3]{\phi_{ki}^{\lambda'}(\beta)}{\hat{H}(\beta)}{\phi_{lj}^{\nu'}(\beta)},
  \label{eq:matrix_transform}
\end{align}
This procedure also applies to $\hat{H}(-\beta)$,
and hence, allows us to directly evaluate the anti-symmetric term $\hat{G}_0(\beta)$
by simply subtracting the matrix elements.
For the small $\beta$,
$\hat{G}_0(\beta)$ is decomposed into two terms
as
\begin{equation}
  \hat{G}_0(\beta)
  = \hat{G}_0^\mathrm{coup}(\beta) + \hat{G}_0^\mathrm{pos}(\beta) + \mathcal{O}(\beta^2),
\end{equation}
where $\hat{G}_0^\mathrm{coup}(\beta)$ depends on $\beta$ from $\cos\gamma$ in the interlayer bonds and $\hat{G}_0^\mathrm{pos}(\beta)$ depends on $\beta$ from the atomic position itself with the non-chiral interlayer coupling elements described by $\sin\gamma$.
The former is expressed as
\begin{equation}
  \braket*[3]{\phi_{2j}^\lambda(0)}{\hat{G}_0^\mathrm{coup}(\beta)}{\phi_{1j}^\nu(0)}
  = \frac{i}{\sqrt{2}}\frac{\rho\beta}{h}
  \begin{pmatrix}
    O          & M \\
    -M^\dagger & O
  \end{pmatrix},
  \label{eq:G0_coup}
\end{equation}
where $\nu$ is taken in order $\{ 0, z, +, - \}$. The $2\times2$ matrix $M$
is defined by
$M_{11} = -M_{12} = W_{sp}^\sigma$ and $M_{21} = -M_{22} = W_{pp}^\sigma-W_{pp}^\pi$.
Thus,
the chirality appears as the imaginary hopping,
which is proportional to the radius $\rho$
and inversely to the interlayer distance $h$.
This feature is consistent with the fact that
$\rho/h$ corresponds to the ratio between the curvature and the torsion in helices.
The latter is expressed as
\begin{equation}
  \braket*[3]{\phi_{2j}^\lambda(0)}{\hat{G}_0^\mathrm{pos}(\beta)}{\phi_{1j}^\nu(0)}
  = \frac{\zeta\rho\beta}{\sqrt{2}}
  \begin{pmatrix}
    0         & 0 & u_1  & u_1^\ast  \\
    0         & 0 & -u_1 & -u_1^\ast \\
    -u_2^\ast & 0 & 0    & 0         \\
    u_2       & 0 & 0    & 0
  \end{pmatrix},
  \label{eq:G0_pos}
\end{equation}
with $u_1 = -ie^{-i(j-1)\alpha}W_{ss}^\sigma/\sqrt{3}$, $u_2 = \sqrt{3}e^{-i(j-1)\alpha}W_{pp}^\pi/2$,
and $\zeta$ being the exponent of the radial component of the orbitals.
The derivation of Eqs.~\eqref{eq:G0_coup} and \eqref{eq:G0_pos} is one of the main results of this work.
These expressions are identified solely from the electronic state of the molecule, and help us to understand how molecular chirality causes various chirality-induced effects.
In this work, we focus on the chirality-induced optical effect in the HHG introduced in the next section and analyze its correspondence with Eqs.~\eqref{eq:G0_coup} and \eqref{eq:G0_pos} in Sec.~\ref{sec:result}.

\section{\label{sec:hhg}Chiral Discriminating High-Harmonic Generation}
\subsection{Three-step model}

Typical spectra and behaviors of HHG for molecular systems can be successfully explained by the three-step model~\cite{corkum1993Plasma,lewenstein1994Theory}:
The electron is first tunnel ionized, which is further accelerated by the laser field.
Finally, the electron recombines with its parent ion emitting the high harmonics.
The intensity of the $n$th harmonics is given by
$I_n = (n\omega)^4\vab*{\bm{\mu}(n\omega)}^2 = (n\omega)^4 \vab*{\int_0^\infty \difc{t}{}\, \bm{\mu}(t) e^{in\omega t}}^2$
with $\bm{\mu}$ being the electric dipole moment.
In the quantum mechanical formulation of the three-step model,
the electron--photon interaction is treated within the electric dipole approximation.
Furthermore, assuming that only the single active electron participates in the HHG process,
the depletion of the ground state can be neglected,
and the nuclear positions are fixed (the frozen nuclear approximation),
the electric dipole moment at time $t$ is represented as
~\cite{lewenstein1994Theory,hasovic2012Strongfield,odzak2016Strongfieldapproximation}
\begin{align}
  \bm{\mu}(t)
   & = \Im \int_0^t \difc{\tau}{}\, C(\tau)
  \bm{d}^\ast[\bm{p}_\mathrm{st}-e\bm{A}(t)]
  e^{-\frac{i}{\hbar} S(\bm{p}_\mathrm{st},t,t-\tau)}
  \notag                                                                     \\
   & \times \bm{d}[\bm{p}_\mathrm{st}-e\bm{A}(t-\tau)] \cdot \bm{E}(t-\tau),
  \label{eq:mu_time}
\end{align}
with $C(\tau)=\frac{2}{\hbar}\ab(\frac{2\pi m_\mathrm{e}\hbar}{\eta+i\tau})^{3/2}$,
the vector potential $\bm{A}$, the electric field $\bm{E}=-\difcp{\bm{A}}{t}$,
the electron charge $e$, and the electron mass $m_\mathrm{e}$.
Here, $\eta$ and $\bm{p}_\mathrm{st}=\int_{t-\tau}^t \difc{t'}{}\,e\bm{A}(t')/\tau$ denote the regularization constant and the stationary momentum, respectively, both of which come from the saddle point approximation in the momentum space integral.
During the acceleration process,
the strong field approximation is invoked, wherein the interaction between the ionized electron and the nuclei is neglected, yielding the propagator described by the action
$S(\bm{p},t,t-\tau)=\int_{t-\tau}^t \difc{t'}{} \{ [\bm{p}-e\bm{A}(t')]^2/2m_\mathrm{e} + I_\mathrm{p} \}$
with $I_\mathrm{p}$ being the ionization potential.
The ionization amplitude is defined as $\bm{d}(\bm{p})=\braket*[3]{\bm{p}}{e\hat{\bm{r}}}{\phi_\mathrm{GS}}$,
where $\hat{\bm{r}}$ is the electron position operator.
In this formulation, the chiral effect is contained in $\bm{d}$ and $\bm{d}^\ast$ through the dependence on $\ket{\phi_\mathrm{GS}(\beta)}$.
In our model,
where the groundstate is given by $\ket*{\phi_\mathrm{GS}(\beta)}=[\ket*{\phi_m(\beta)}+\ket*{\phi_{-m}(\beta)}]/\sqrt{2}$,
$\bm{\mu}(t)$ is a sum of the IRREP-resolved electric dipoles, $\bm{\mu}(t) = [\bm{\mu}_m(t) + \bm{\mu}_{-m}(t)]/2$,
where $\bm{\mu}_m(t)$ is obtained by replacing $\ket{\phi_{\mathrm{GS}}(\beta)}$ in Eq.~\eqref{eq:mu_time} with $\ket*{\phi_m(\beta)}$, because the ionized electron returns to its original state or orbital,
and hence, the mixing of the states with the different $m$ can be neglected.

\subsection{Pseudoscalar property of HHG signal}

The chiral discrimination is quantified by the Kuhn $g$ factor~\cite{kuhn1930Physical},
which is generalized to the HHG signals as defined below.
To prove that the Kuhn $g$ factor is a pseudoscalar quantity,
we assume that the molecular rotational axis is fixed and parallel with the propagation direction of the light beams.
This assumption affects the signals quantitatively, but is not the prerequisite for the pseudoscalar nature as exemplified by the numerical~\cite{ayuso2018Chiral} and experimental~\cite{cireasa2015Probinga,harada2018Circular} studies for randomly oriented molecules.
Then, the vector potential for the bicircular field is given by
\begin{equation}
  \bm{A}(t)
  = A_+ \bm{e}_+ e^{-ir_+\omega t}
  + A_- \bm{e}_- e^{-ir_-\omega t}
  + \mathrm{c.c.},
  \label{eq:light_field}
\end{equation}
where $\bm{e}_\pm=\frac{1}{\sqrt{2}}(1, \pm i, 0)^\top$ are the polarization vectors
and $A_\pm$ is the corresponding amplitudes
with the fundamental frequency $\omega$ and $r_\pm \in \mathbb{N}$.
This is transformed under the reflection operation to
$\bar{\bm{A}}(t)= \hat{\sigma}\bm{A}(t)$ obtained by interchanging the polarization vectors, $\bm{e}_\pm \to \bm{e}_\mp$.
In the cases of  $A_+ \ne A_-$ or $r_+ \ne r_-$, the inequality $\bar{\bm{A}}(t) \ne \bm{A}(t)$ holds,
and thus, the bicircular field is chiral.
The reflection operation also acts as
$\hat{\sigma}\ket*{\phi_m(\beta)} \equiv \ket*{\phi_{-m}(-\beta)}$,
which straightforwardly leads to
$\bm{\mu}(n\omega,\bm{A},\beta) = \hat{\sigma}\bm{\mu}(n\omega,\bar{\bm{A}},-\beta)$,
and thus, $I_n(\bm{A},\beta) = I_n(\bar{\bm{A}},-\beta)$.
Now, it is clear that the Kuhn $g$ factor
\begin{equation}
  g(\beta)
  \coloneqq 2 \frac{I_n(\bm{A},\beta)-I_n(\bar{\bm{A}},\beta)}{I_n(\bm{A},\beta)+I_n(\bar{\bm{A}},\beta)},
  \label{eq:gfactor}
\end{equation}
is pseudoscalar, $g(\beta) = -g(-\beta)$, for all harmonic orders $n$.
We also introduce the unnormalized $g$ factor, which we call $\tilde{g}$ factor,
as $\tilde{g}(\beta) \coloneqq I_n(\bm{A},\beta)-I_n(\bar{\bm{A}},\beta)$ to discuss the correspondence between the molecular chirality and the HHG signal.
The $\tilde{g}$ factor is also the pseudoscalar.

\section{\label{sec:result}Results}

To calculate the electric dipole moment Eq.~\eqref{eq:mu_time},
we performed the numerical integration using the Gauss--Kronrod quadrature method~\cite{WilliamsModern}
with the quad precision arithmetics.
Throughout the calculation,
we set to $\omega A_+ = \qty[per-mode=symbol]{3d10}{\volt\per\metre}$,
$\omega A_- = \qty[per-mode=symbol]{4d10}{\volt\per\metre}$, $r_+ = 1$, $r_- = 2$, and $\omega = \qty{1}{\eV}$ for the laser field,
and $\rho=\qty{0.256}{\nm}$, $V = \qty{0.5}{\eV}$, $W = \qty{0.2}{\eV}$, and $I_\mathrm{p} = \qty{14}{\eV}$ for the chiral molecule.
Hereafter, we focus on the $N=3$ with $m=\pm1$ case only,
but the qualitative consistency was also checked for the $N=4$ with $m=\pm1$ and the $N=6$ with $m=\pm1$ and $\pm2$ cases.

\begin{figure}
  \centering
  \includegraphics[width=81mm]{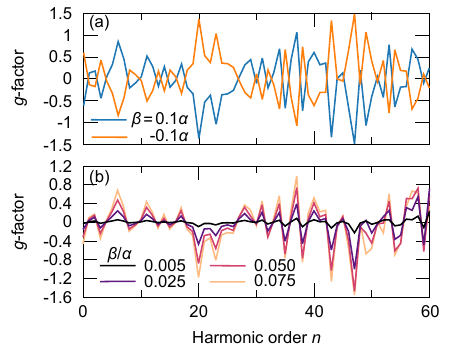}
  \caption{(a) Calculated $g$ factors for $\beta=0.1\,\alpha$ (blue line) and $\beta=-0.1\,\alpha$ (orange line).
    Each enantiomer is interconverted to the other by inverting $\beta$. (b) Twist angle, $\beta$, dependence of the $g$-factor. The interlayer distance is fixed to $h=\qty{0.6}{\nm}$.}
  \label{fig:pseudoscalar}
\end{figure}

\begin{figure}
  \centering
  \includegraphics[width=81mm]{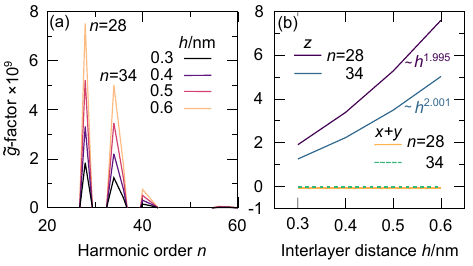}
  \caption{(a) Calculated $\tilde{g}$ factors for various values of the interlayer distance, $h$.
    (b) The $\tilde{g}$ factor decomposition into the $x+y$ component, $\vab*{\mu_x(n\omega)}^2+\vab*{\mu_y(n\omega)}^2$, and the $z$ component, $\vab*{\mu_z(n\omega)}^2$, for the harmonic orders $n=28$ and $34$. Twist angle is fixed to $\beta = 0.005\,\alpha$.}
  \label{fig:h_dependence}
\end{figure}

Figure~\ref{fig:pseudoscalar}(a) depicts the $g$ factor for each enantiomer of our model chiral molecule.
The pseudoscalar property is confirmed for all harmonic orders $n$,
because the $g$ factor is reversed by inverting the twist angle, $\beta$.
The $g$ factor has a value in $-2 \le g \le 2$ and its increasing is advantageous for the chiral discrimination.
In Fig.~\ref{fig:pseudoscalar}(a), this value can exceeds \num{1.0}, which is much larger than typical values of the CD.
This is because the CD is contributed from the magnetic dipole moment in the light--matter interaction, the magnitude of which is sufficiently smaller than that of the electric dipole moment.
At this point, we note that the harmonic intensity itself becomes smaller for the higher harmonic orders and the largeness of the $g$ factor comes from the denominator (normalization) of Eq.~\eqref{eq:gfactor}. Thus, the effectiveness of the HHG for chiral discrimination refers to using the $g$ factor for the harmonic orders measurable with high accuracy.

Next, we examine the correspondence between the molecular chirality of Eqs.~\eqref{eq:G0_coup}, \eqref{eq:G0_pos} and the $g$-factor.
Hereafter, we focus on the small $\beta$ region,
where the Hamiltonian is approximated as
$\hat{H}(\beta) \approx \hat{H}(0) + \hat{G}_0^\mathrm{coup}(\beta) + \hat{G}_0^\mathrm{pos}(\beta)$.
In Fig.~\ref{fig:pseudoscalar}(b), we present the $g$ factors for the various values of $\beta$,
where the $g$ factor monotonically increases as $\beta$ is increased.
Because the twist angle does not change the molecular size, which is the another element to affect the signal,
this feature can be attributed to the linear $\beta$ dependence of $\hat{G}_0^\mathrm{coup}(\beta)$ and $\hat{G}_0^\mathrm{pos}(\beta)$, and thus, the increase of the molecular chirality.

To analyze the correspondence without consideration of the effects from $\hat{G}_0^\mathrm{pos}(\beta)$,
we consider the $h$ dependence of the chiral discriminating signals,
because $\hat{G}_0^\mathrm{pos}(\beta)$ has no $h$-dependence.
To remove further the contribution from the symmetric part of the Hamiltonian
of the denominator in Eq.~\eqref{eq:light_field},
we present the $\tilde{g}$ factors for various values of $h$ in Fig.~\ref{fig:h_dependence}(a),
where the peak intensities are a monotonic increasing function of $h$.
For the peaks of $n=28$ and $34$,
we decompose this into $\vab*{\mu_x(n\omega)}^2+\vab*{\mu_y(n\omega)}^2$ and $\vab*{\mu_z(n\omega)}^2$ depicted in Fig.~\ref{fig:h_dependence}(b),
where the $x+y$ component is negligibly small, and thus, the $z$ component dominantly contributes to the $\tilde{g}$ factor.
We further found that this $z$ component scales quadratically with $h$.
To unveil the origin of this, the ionization amplitude $\bm{d}(\bm{p})$ is expanded with the basis states as
\begin{equation}
  \bm{d}(\bm{p})
  = \sum_{kj\nu} b_{kj}^\nu e^{-i\bm{p}\cdot\bm{R}_{kj}/\hbar}
  \ab( \bm{R}_{kj} + i\hbar \difcp*{}{\bm{p}}) \tilde{\phi}^\nu(\bm{p}),
  \label{eq:d_mol}
\end{equation}
where $b_{kj}^\nu$ is the linear coefficient for the basis $\ket*{\phi_{kj}^\nu}$ and $\tilde{\phi}^\nu(\bm{p})$ is the momentum representation of $\phi^\nu$.
Eqs.~\eqref{eq:mu_time} and \eqref{eq:d_mol} shows that the quadratic dependence is originated from the $z$ component, $(\bm{R}_{kj})_z=(k-1)h$, of the first term of Eq.~\eqref{eq:d_mol} for the recombination process, $\bm{d}^\ast$,
and not the ionization process.
This is because the ionization amplitude emerges as the form $\bm{d}\cdot\bm{E}$
and does not depend on this term due to no $z$ component of $\bm{E}(\tau)$.
Thus, the chirality-induced optical process in the HHG can be interpreted as the conversion from the transverse dipole moment generated in the ionization process to the longitudinal one at the recombination.
Because increasing the interlayer distance, $h$, decreases $\hat{G}_0^\mathrm{coup}(\beta)$,
simultaneously increasing the molecular size,
this result also indicates that the enhancement of the electric dipole moment by enlarging the molecular size contributes dominantly to the $\tilde{g}$ factor rather than increasing the molecular chirality, $\hat{G}_0^\mathrm{coup}(\beta)$.
This may be because the $z$-component of Eq.~\eqref{eq:d_mol} is proportional to $h$ and unbounded,
whereas the chiral contributions from $b_{kj}^\nu$ and $\bm{R}_{kj}$ are bounded by the normalization condition of the wave function and the helical angle $\gamma$ appeared as $\sin\gamma$ and $\cos\gamma$, respectively.
What should be noted here is that this $z$ component is contributed only from the $2\mathrm{p}_z$ orbitals, which are involved only in the chiral cases.
Indeed, in the achiral case with $\beta=0$, the $2\mathrm{p}_z$ orbital belong to the different IRREP from the other orbitals, and thus, they are orthogonal.

\section{\label{sec:conclu}Concluding Remarks}

In this study, we have developed the analytical method that extracts the anti-symmetric part of the Hamiltonian
that describes the chirality.
Our approach is based on the translation of the orbital center of each electronic basis
from the achiral system to the chiral system.
This allows us to expand the chiral Hamiltonians with the achiral basis states,
enabling the direct subtraction of the matrix elements of the different chirality.
As a demonstration of this method,
we consider the model chiral molecule that consists of two helically stacked $N$-sided regular polygons
and investigate the correspondence between the chiral part of our model and the Kuhn $g$ factor in HHG.
Numerical results shows that the $g$ factor is the pseudoscalar quantity reflecting the chirality of the molecule.
Calculated values are much larger than the typical values of the CD
and slightly larger than that reported in Ref.~\cite{harada2018Circular} of the gaseous molecules.
Extension of our result to the randomly oriented molecules may give the quantitatively consistent result with the experiments.
Numerical results also reveal that the enhancement of the electric dipole moment dominantly contributes to increasing the difference of the harmonic intensities even when this enhancement is accompanied by reducing the molecular chirality.
Therefore, we speculate that the above criteria may be generalized to interconversions between polar and axial quantities: This interconversion is induced by the chirality and works efficiently for the strong chiral systems. However, to obtain the large axial quantity from the polar one, increasing capacity to generate the axial quantity can be effective (and vice versa).
As the CD monotonically increases with the molecular length~\cite{mishra2020LengthDependent}, we expect that this speculation is valid to various optical processes.
Furthermore, this may also be relevant to the effects induced by the metal--molecule interaction with or without the charge injection in the chiral systems, especially the CISS, wherein the interrelation between the charge redistribution and the formation of the anti-parallel spin pair~\cite{kumar2017Chiralityinduced,bendor2017Magnetization,nakajima2023Giant} is hypothesized as one of the keys to understand the CISS mechanism~\cite{fransson2021Charge,bloom2024Chiral}.

Our method developed in this paper is applicable to real chiral molecules, helices, and crystals with the values of their orbital overlap integrals.
Generalization to the systems with the spin--orbit, electron--phonon, and electron--photon interactions will be of great interest
in elucidating the microscopic insight into the AM exchange among them~\cite{li2022Energy,bian2024angular,maslov2024Theory},
which is left for future studies.

\begin{acknowledgments}
  This work was supported by the JSPS KAKENHI Grants No. 21H05019, No. 22K04863, and No. 22H05132.
\end{acknowledgments}

\appendix
\section{\label{sec:tb_detail}Construction of tight-binding Hamiltonian}

In this Appendix,
the tight-binding Hamiltonian of the model chiral molecule introduced in Sec.~\ref{sec:model}
is constructed using the Slater--Koster method
following Refs.~\cite{harrison1989electronic,varela2016Effective}.
For notational simplicity, we omit $\beta$ to specify the chirality from the description.

The matrix element of the Hamiltonian is written as
\begin{equation}
  H_{ki,lj}^{\lambda\nu}
  \coloneqq \braket*[3]{\phi_{ki}^\lambda}{\hat{H}}{\phi_{lj}^\nu}
  = E_{ki,lj}^{\lambda\nu} \exp\ab(i\xi_{ki,lj}^{\lambda\nu}),
\end{equation}
where $E_{ki,lj}^{\lambda\nu}$ is the overlap integral
and $\xi_{ki,lj}^{\lambda\nu}$ is the phase factor that is determined to maintain the molecular symmetry.
Let $\hat{\bm{n}}(\nu_{ki})$ denote the unit vector along the orbital $\ket*{\phi_{ki}^\nu}$.
Then, the overlap integral can be expressed as
\begin{align}
  E_{ki,lj}^{\lambda\nu}
   & = \aab*{\hat{\bm{n}}(\lambda_{ki}), \hat{\bm{n}}(\nu_{lj})} V_{\lambda\nu}^\pi
  \notag                                                                                                           \\
   & + \aab*{\hat{\bm{R}}_{lj,ki}, \hat{\bm{n}}(\lambda_{ki})} \aab*{\hat{\bm{R}}_{lj,ki}, \hat{\bm{n}}(\nu_{lj})}
  \ab(V_{\lambda\nu}^\sigma-V_{\lambda\nu}^\pi),
  \label{eq:SK_overlap}
\end{align}
where $V_{\lambda\nu}^\sigma$ and $V_{\lambda\nu}^\pi$ are the Slater--Koster overlaps
and $\bm{R}_{lj,ki} = \bm{R}_{lj}-\bm{R}_{ki}$ is the bond vector.
The normalized bond vector, $\hat{\bm{R}}_{lj,ki} \equiv \bm{R}_{lj,ki}/\vab*{\bm{R}_{lj,ki}}$,
is given by
\begin{align}
  \hat{\bm{R}}_{lj,ki}
   & = -\sin(\theta_{lj}+\theta_{ki})\cos\gamma_{lj,ki} \bm{e}_x
  \notag                                                              \\
   & \quad + \cos(\theta_{lj}+\theta_{ki})\cos\gamma_{lj,ki} \bm{e}_y
  \notag                                                              \\
   & \quad + \sin\gamma_{lj,ki} \bm{e}_z
\end{align}
with $\theta_{ki} = [(k-1)\beta+(i-1)\alpha]/2$, $\theta_{lj,ki}=\theta_{lj}-\theta_{ki}$,
and $\gamma_{lj,ki} = \arctan[(l-k)h / 2\rho\sin\theta_{lj,ki}]$.
With these notations,
$\hat{\bm{n}}(\nu_{ki})$ is written as
$\hat{\bm{n}}(s_{ki}) = \hat{\bm{R}}_{lj,ki}$,
$\hat{\bm{n}}(s_{lj}) = -\hat{\bm{R}}_{lj,ki}$,
$\hat{\bm{n}}(x_{ki}) = \cos2\theta_{ki} \bm{e}_x + \sin2\theta_{ki} \bm{e}_y$,
$\hat{\bm{n}}(y_{ki}) = -\sin2\theta_{ki} \bm{e}_x + \cos2\theta_{ki} \bm{e}_y$,
and $\hat{\bm{n}}(z_{ki}) = \bm{e}_z$.
Hence, we can obtain the analytical expressions of the overlap integrals, $E_{ki,lj}^{\lambda\nu}$,
by directly evaluating Eq.~\eqref{eq:SK_overlap}.
The details of the calculation are given in Refs.~\cite{varela2016Effective}.
Finally, we can obtain the matrix elements for the intralayer bond as
\begin{align}
  H_{kj,kj\pm1}^{ss}
   & = -V_{ss}^\sigma
  \\
  H_{kjkj\pm1}^{zz}
   & = V_{pp}^\pi
  \\
  H_{kj,kj\pm1}^{zs}
   & = H_{kj,kj\pm1}^{sz}
  = 0
  \\
  H_{kj,kj\pm1}^{s\eta}
   & = -H_{kj,kj\pm1}^{\eta s}
  = \pm \frac{i\eta}{\sqrt{2}} e^{\pm i \eta \pi/2} V_{sp}^\sigma
  \\
  H_{kj,kj\pm1}^{z\eta}
   & = H_{kj,kj\pm1}^{\eta z}
  = 0
  \\
  H_{kj,kj\pm1}^{\eta\eta}
   & = \frac{1}{2} e^{\mp i\eta \alpha} \ab( V_{pp}^\sigma + V_{pp}^\pi )
  \\
  H_{kj,kj\pm1}^{\eta\bar{\eta}}
   & = - \frac{1}{2} \ab( V_{pp}^\sigma - V_{pp}^\pi )
\end{align}
and for the interlayer bond as
\begin{align}
  H_{kj,k\pm1j}^{ss}
   & = -W_{ss}^\sigma
  \\
  H_{kj,k\pm1j}^{zz}
   & = W_{pp}^\pi
  + \sin^2\gamma \ab(W_{pp}^\sigma-W_{pp}^\pi)
  \\
  H_{kj,k\pm1j}^{zs}
   & = H_{kj,k\pm1j}^{sz}
  = \pm \sin\gamma W_{sp}^\sigma
  \\
  H_{kj,k\pm1j}^{s\eta}
   & = \ab( H_{kj,k\pm1j}^{\eta s} )^\ast
  = \pm \frac{i\eta}{\sqrt{2}} e^{i(j-1)\eta\alpha} \cos\gamma W_{sp}^\sigma
  \\
  H_{kj,k\pm1j}^{z\eta}
   & = \ab( H_{kj,k\pm1j}^{\eta z} )^\ast
  \notag                                                                                          \\
   & = \frac{i\eta}{2\sqrt{2}} e^{i(j-1)\eta\alpha} \sin2\gamma \ab( W_{pp}^\sigma - W_{pp}^\pi )
  \\
  H_{kj,k\pm1j}^{\eta\eta}
   & = W_{pp}^\pi
  + \frac{1}{2} \cos^2\gamma \ab( W_{pp}^\sigma - W_{pp}^\pi )
  \\
  H_{kj,k\pm1j}^{\eta\bar{\eta}}
   & = -\frac{1}{2} \cos^2\gamma \ab( W_{pp}^\sigma - W_{pp}^\pi ).
\end{align}
In the above, the unknown parameters are the Slater--Koster overlap constants.
To determine these, we take the carbon atom as a reference~\cite{harrison1989electronic,varela2016Effective} and set to
$V_{ss}^\sigma = -1.40 \,V$, $V_{sp}^\sigma =  1.84 \,V$,
$V_{pp}^\sigma =  3.24 \,V$, $V_{pp}^\pi = -0.81 \,V$,
and
$W_{ss}^\sigma = -1.40 \,W$, $W_{sp}^\sigma =  1.84 \,W$,
$W_{pp}^\sigma =  3.24 \,W$, $W_{pp}^\pi = -0.81 \,W$.

\section{\label{sec:g0_detail}Extraction of the anti-symmetric part of the Hamiltonian}

In this Appendix,
we present a method to evaluate the anti-symmetric part of the Hamilonian,
and identify the key parameters describing the chirality of the system.

The Hamiltonian is separated formally into the symmetric and anti-symmetric terms
with respect to the reflection operation as
\begin{equation}
  \hat{H}(\beta)
  = \hat{Q}_0(\beta) + \hat{G}_0(\beta),
\end{equation}
where $\hat{Q}_0(\beta) = [\hat{H}(\beta)+\hat{H}(-\beta)]/2$
and $\hat{G}_0(\beta) = [\hat{H}(\beta)-\hat{H}(-\beta)]/2$
satisfy $\hat{\sigma}\hat{Q}_0(\beta)\hat{\sigma}^\dagger = \hat{Q}_0(\beta)$
and $\hat{\sigma}\hat{G}_0(\beta)\hat{\sigma}^\dagger = -\hat{G}_0(\beta)$, respectively.
The anti-symmetric term $\hat{G}_0$ is defined as the subtraction of the Hamiltonians,
each of which has the different chirality and is expanded in terms of the chirality-dependent electronic basis states.
Because the vanishing twist angle, $\beta=0$, corresponds to the achiral system,
and using the fact that the chirality enters as the atomic positions in the orbital center,
each basis is related to the achiral basis as
\begin{align}
  \phi^\nu[\bm{r}-\bm{R}_{kj}(0)]
   & = \exp\ab(-\frac{i}{\hbar}\bm{\hat{p}}\cdot[\bm{R}_{kj}(0)-\bm{R}_{kj}(\beta)] )
  \notag                                                                              \\
   & \times \phi^\nu[\bm{r}-\bm{R}_{kj}(\beta)]
  \notag                                                                              \\
   & = \hat{T}_{kj}(\beta) \phi^\nu[\bm{r}-\bm{R}_{kj}(\beta)],
\end{align}
where $\hat{\bm{p}}=-i\hbar\grad$ is the momentum operator.
The exponent is expressed as
\begin{align}
   & -\frac{i}{\hbar}\hat{\bm{p}} \cdot [\bm{R}_{kj}(0) - \bm{R}_{kj}(\beta)]
  \notag                                                                      \\
   & = i \rho \sin\ab(\frac{k-1}{2}\beta) \ab[
    e^{i\kappa_{kj}} \ab( \partial_x - i \partial_y )
    - e^{-i\kappa_{kj}} \ab( \partial_x + i \partial_y ) ].
\end{align}
with $\kappa_{kj} = (k-1)\beta/2+(j-1)\alpha$.
Because the nuclei are fixed, the differential operator for $\bm{r}$
is equivalent to that for $\bm{r}' = \bm{r}-\bm{R}_{kj}(\beta) = (x',y',z')^\top$,
$\partial_{a} = \partial_{a'}$ ($a=x,y,z$).
By expressing the orbitals as $\phi^\nu(\bm{r}') = R_n(r') Y_l^m(\theta',\varphi')$,
where $R_n$ and $Y_l^m$ are the radial distributions and the spherical harmonics
with $n, l, m$ being the quantum number corresponding to the orbital $\nu$,
the differential operator acts as
\begin{align}
  \ab( \partial_{x'} \pm i \partial_{y'} ) \phi^\nu
   & = e^{\pm i\varphi'} \left[ \sin\theta'  Y_l^m \difcp{R_n}{r'} \right.
  \notag                                                                   \\
   & \left. + \frac{R_n}{r'} \ab( \cos\theta' \difcp{}{\theta'}
    \pm i \frac{1}{\sin\theta'} \difcp{}{\varphi'} ) Y_l^m \right].
\end{align}
Assuming that the Hamiltonian is spanned by the $2\mathrm{s}$, $2\mathrm{p}_z$, and $2\mathrm{p}_\pm$ orbitals,
and hence, the contribution from the other is negligibly small,
the action of the differential operator becomes
\begin{align}
  \ab( \partial_x \pm i \partial_y ) \phi^0(\bm{r}')
   & = - \sqrt{\frac{2}{3}} \zeta \phi^\pm(\bm{r}')
  \\
  \ab( \partial_x \pm i \partial_y ) \phi^z(\bm{r}')
   & = 0                                            \\
  \ab( \partial_x \pm i \partial_y ) \phi^\pm(\bm{r}')
   & = 0
  \\
  \ab( \partial_x \pm i \partial_y ) \phi^\mp(\bm{r}')
   & = - \sqrt{\frac{3}{2}} \zeta \phi^0(\bm{r}'),
  \label{eq:diff_xy_orbital}
\end{align}
with $\zeta$ being the exponent of the radial distribution, $R_n(r) \propto r^{n-1} e^{-\zeta r}$.

What is necessary is to represent the action of $\hat{T}_{kj}(\beta)$
on the orbital space $\{ \phi^\nu(\bm{r}')\}_\nu$,
namely,
to determine the coefficients that satisfies
$\hat{T}_{kj}(\beta)\phi^\nu(\bm{r}') = \sum_{\nu'} g_{kj}^{\nu\nu'}(\beta) \phi^{\nu'}(\bm{r}')$
with $g_{kj}^{\nu\nu'}(\beta) \in \mathbb{C}$.
For that purpose,
the eigenstates of the operator
$i[e^{i\kappa_{kj}}(\partial_x-i\partial_y) - e^{-i\kappa_{kj}}(\partial_x+i\partial_y)]$
as the linear coupling of $\bm{\phi}' = [\phi^0(\bm{r}'), \phi^+(\bm{r}'), \phi^-(\bm{r}')]^\top$
needs to be constructed.
The orbital $\phi^z(\bm{r}')$ does not couple to the other orbital and itself forms the eigenstate.
From Eq.~\eqref{eq:diff_xy_orbital}, the action of the above operator is written as
\begin{align}
   & i[e^{i\kappa_{kj}}(\partial_x-i\partial_y) - e^{-i\kappa_{kj}}(\partial_x+i\partial_y)] \bm{\phi}'
  \notag                                                                                                \\
   & = -\sqrt{\frac{2}{3}} \zeta
  \begin{pmatrix}
    0                               & -ie^{-i\kappa_{kj}} & ie^{i\kappa_{kj}} \\
    \frac{3}{2}ie^{i\kappa_{kj}}    & 0                   & 0                 \\
    -\frac{3}{2} ie^{-i\kappa_{kj}} & 0                   & 0
  \end{pmatrix}
  \bm{\phi}'.
  \label{eq:diff_matrix}
\end{align}
Hence, we have to diagonalize the matrix in the right-hand side of Eq.~\eqref{eq:diff_matrix}.
This can be straightforwardly done, resulting to the eigenvalues $\epsilon_0 = 0$ and $\epsilon_\pm = \mp \sqrt{2} \zeta$
with the corresponding eigenstates
\begin{align}
  \psi_0
   & = \frac{1}{\sqrt{2}} \ab( e^{i\kappa_{kj}} \phi^+ + e^{-i\kappa_{kj}} \phi^- ),
  \\
  \psi_\pm
   & = \mp i \sqrt{\frac{2}{5}} \phi^0
  + \sqrt{\frac{3}{10}} \ab( e^{i\kappa_{kj}} \phi^+ - e^{-i\kappa_{kj}}\phi^- ),
\end{align}
respectively.
Inversely, each orbital can be expressed in terms of the eigenstates as
\begin{align}
  \phi^0
   & = \sqrt{\frac{5}{8}} i \ab( \psi_+ - \psi_- )
  \\
  \phi^\pm
   & = e^{\mp i\kappa_{kj}} \ab[ \frac{1}{\sqrt{2}} \psi_0 \pm \sqrt{\frac{5}{24}} \ab( \psi_+ + \psi_- ) ].
\end{align}

Using these results,
the action of $\hat{T}_{kj}(\beta)$ on the orbitals $\bm{\phi}=[\phi^0(\bm{r}'), \phi^z(\bm{r}'), \phi^+(\bm{r}'), \phi^-(\bm{r}')]^\top$ is calculated
and summarized as $\hat{T}_{kj}(\beta)\bm{\phi} = g_{kj}(\beta)\bm{\phi}$ with
\begin{align}
   & g_{kj}(\beta)
  \notag                                                                                                                                                                                        \\
   & = \begin{pmatrix}
         c_k                                      & 0 & -\frac{\sqrt{3}}{2} s_k e^{i\kappa_{kj}} & \frac{\sqrt{3}}{2} s_k e^{-i\kappa_{kj}} \\
         0                                        & 1 & 0                                        & 0                                        \\
         \frac{i}{\sqrt{3}} s_k e^{-i\kappa_{kj}} & 0 & \frac{1+c_k}{2}                          & \frac{1-c_k}{2} e^{-i2\kappa_{kj}}       \\
         -\frac{i}{\sqrt{3}} s_k e^{i\kappa_{kj}} & 0 & \frac{1-c_k}{2} e^{i2\kappa_{kj}}        & \frac{1+c_k}{2}
       \end{pmatrix}
  \label{eq:g_matrix}
\end{align}
with $s_k = \sinh\tau_k$, $c_k = \cosh\tau_k$,
and $\tau_k = \sqrt{2}\zeta\rho\sin(\frac{k-1}{2}\beta)$.

Now, let us calculate the anti-symmetric part of the Hamiltonian,
$\hat{G}_0(\beta) = [\hat{H}(\beta)-\hat{H}(-\beta)]/2$ expanded in terms of the achiral basis states.
The matrix element is given by
\begin{align}
   & \braket*[3]{\phi_{ki}^\lambda(0)}{\hat{G}_0(\beta)}{\phi_{lj}^\nu(0)}
  \notag                                                                       \\
   & = \frac{1}{2}\sum_{\lambda'\nu'}
  \left[ [g_{ki}^{\lambda\lambda'}(\beta)]^\ast g_{lj}^{\nu\nu'}(\beta)
  \braket*[3]{\phi_{ki}^{\lambda'}(\beta)}{\hat{H}(\beta)}{\phi_{lj}^{\nu'}(\beta)}
  \right.
  \notag                                                                       \\
   & \left. - [g_{ki}^{\lambda\lambda'}(-\beta)]^\ast g_{lj}^{\nu\nu'}(-\beta)
  \braket*[3]{\phi_{ki}^{\lambda'}(-\beta)}{\hat{H}(-\beta)}{\phi_{lj}^{\nu'}(-\beta)}
  \right].
\end{align}
Using Eq.~\eqref{eq:g_matrix},
the right-hand side of the above equation can be calculated,
which has the matrix form in the orbital space
\begin{equation}
  \braket*[3]{\phi_{2j}^\lambda(0)}{\hat{G}_0(\beta)}{\phi_{1j}^\nu(0)}
  = \begin{pmatrix}
    f_1                 & 0        & f_3           & f_3^\ast      \\
    f_2                 & 0        & f_4           & f_4^\ast      \\
    f_5 + f_6           & f_7      & f_8           & -f_8^\ast+f_9 \\
    f_5^\ast - f_6^\ast & f_7^\ast & -f_8+f_9^\ast & f_8^\ast      \\
  \end{pmatrix},
\end{equation}
where its elements are defined as
\begin{align}
  f_1
   & = -i\sqrt{\frac{3}{2}} \Im(a_1) W_{sp}^\sigma
  \\
  f_2
   & = -i\sqrt{\frac{3}{2}} \Im(a_1) \sin\gamma (W_{pp}^\sigma-W_{pp}^\pi)
  \\
  f_3
   & = -\frac{i}{\sqrt{3}} a_2 W_{ss}^\sigma + \frac{i}{\sqrt{2}} a_3 W_{sp}^\sigma
  \\
  f_4
   & = \frac{i}{\sqrt{3}} a_2 \sin\gamma W_{ss}^\sigma + \frac{i}{\sqrt{2}} a_3 \sin\gamma (W_{pp}^\sigma-W_{pp}^\pi)
  \\
  f_5
   & = -\frac{i}{\sqrt{2}} \cosh\tau \cos\gamma W_{sp}^\sigma
  \\
  f_6
   & = -\frac{\sqrt{3}}{2} \ab( a_2^\ast W_{pp}^\pi + \Re(a_2) \cos^2\gamma (W_{pp}^\sigma-W_{pp}^\pi) )
  \\
  f_7
   & = -\frac{i}{\sqrt{2}} \sin\gamma \cos\gamma (W_{pp}^\sigma-W_{pp}^\pi)
  \\
  f_8
   & = \frac{i}{\sqrt{6}} a_1 W_{sp}^\sigma
  - \frac{i}{2} a_4 \cos^2\gamma (W_{pp}^\sigma-W_{pp}^\pi)
  \\
  f_9
   & = -i a_4^\ast W_{pp}^\pi
\end{align}
with
\begin{align}
  a_1
   & = e^{-i(j-1)\alpha} \sinh\tau \sin\frac{\beta}{2} \cos\gamma
  \\
  a_2
   & = e^{-i(j-1)\alpha} \sinh\tau \cos\frac{\beta}{2}
  \\
  a_3
   & = \cosh^2\frac{\tau}{2} + e^{-2i(j-1)\alpha} \sinh^2\frac{\tau}{2}\cos\beta
  \\
  a_4
   & = e^{-2i(j-1)\alpha} \sinh^2\frac{\tau}{2} \sin\beta
\end{align}
and $\tau = \tau_2 = \sqrt{2}\zeta \rho \sin(\beta/2)$.
In the above elements,
we can distinguish two origins of $\beta$-dependence,
which are the atomic position through the dependence on $j$
and the interlayer bond through the dependence on the helical angle $\gamma$.
These two contributions are decouped in the first order component of $\beta$.
For the sufficiently small $\beta$ region,
the elements from the former and the latter origins are expressed as $f_i^\mathrm{pos}$ and $f_i^\mathrm{coup}$, respectively and are written as
$f_1^\mathrm{pos} = f_2^\mathrm{pos} = f_5^\mathrm{pos} = f_7^\mathrm{pos} = f_8^\mathrm{pos} = f_9^\mathrm{pos} = 0$,
\begin{align}
  f_3^\mathrm{pos}
   & = -\frac{i}{\sqrt{6}} \zeta \rho \beta e^{-i(j-1)\alpha} W_{ss}^\sigma,
  \\
  f_4^\mathrm{pos}
   & = \frac{i}{\sqrt{6}} \zeta \rho \beta e^{-i(j-1)\alpha} W_{ss}^\sigma,
  \\
  f_6^\mathrm{pos}
   & = -\sqrt{\frac{3}{8}} \zeta \rho \beta e^{i(j-1)\alpha}W_{pp}^\pi,
\end{align}
$f_1^\mathrm{coup} = f_2^\mathrm{coup} = f_6^\mathrm{coup} = f_8^\mathrm{coup} = f_9^\mathrm{coup} = 0$,
\begin{align}
  f_3^\mathrm{coup}
   & = \frac{i}{\sqrt{2}} \frac{\rho}{h} \beta W_{sp}^\sigma,
  \\
  f_4^\mathrm{coup}
   & = \frac{i}{\sqrt{2}} \frac{\rho}{h} \beta (W_{pp}^\sigma-W_{pp}^\pi),
  \\
  f_5^\mathrm{coup}
   & = -\frac{i}{\sqrt{2}} \frac{\rho}{h} \beta W_{sp}^\sigma,
  \\
  f_7^\mathrm{coup}
   & = -\frac{i}{\sqrt{2}} \frac{\rho}{h} \beta (W_{pp}^\sigma-W_{pp}^\pi).
\end{align}

\bibliography{ref_chiral_hhg}

\end{document}